\documentclass[prd,nofootinbib,twocolumn]{revtex4}

\usepackage{amssymb}
\usepackage{amsmath}
\usepackage{graphicx,subfigure}
\usepackage{longtable}
\usepackage{verbatim}
\usepackage{amsfonts}

\allowdisplaybreaks

\newcommand{\be}{\begin{equation}}
\newcommand{\ee}{\end{equation}}
\newcommand{\bea}{\begin{eqnarray}}
\newcommand{\eea}{\end{eqnarray}}
\newcommand{\bv}{\left(\begin{array}{c}}
\newcommand{\ev}{\end{array}\right)}

\newcommand{\refEqws}[1]{Eq.~(\ref{#1})}

\begin{document}
\begin{flushright}
{\footnotesize DO-TH 12/43}
\end{flushright}

\vspace*{-30mm}

\title{
A robust limit for the electric dipole moment of the electron}
\author{Martin Jung}
\email{martin2.jung@tu-dortmund.de}
\affiliation{Institut f\"ur Physik, Technische Universit\"at Dortmund, D-44221 
Dortmund, Germany}
\vspace*{1cm}
\begin{abstract}
Electric dipole moments constitute a competitive method to search for new physics, being particularly sensitive to new $CP$-violating phases. Given the experimental and theoretical progress in this field and more generally in particle physics, the necessity for more reliable bounds than the ones usually employed emerges. We therefore propose an improved extraction of the electric dipole moment of the electron and the relevant coefficient of the electron-nucleon coupling, taking into account theoretical uncertainties and possible cancellations, to be used in model-dependent analyses. Specifically, we obtain at $95\%$~C.L. $|d_e|\leq0.14\times 10^{-26}e\,\rm{cm}$ with present data, which is very similar to the bound typically quoted from the YbF molecule, but obtained in a more conservative manner.  We examine furthermore in detail the prospects for improvements and derive upper limits for the dipole moments of several paramagnetic systems presently under investigation, i.e. Cesium, Rubidium and Francium.  
\end{abstract}

\maketitle

\section{Introduction}
Despite the tremendous success of the Standard Model (SM), the arguments for the necessity of an extension are compelling. Specifically, Sakharov's conditions \cite{Sakharov:1967dj} require the presence of additional sources for $CP$ violation with respect to the SM, given the observed baryon asymmetry of the universe. Assuming $CPT$ invariance, electric dipole moments (EDMs) are highly sensitive probes for new $CP$-violating phases. This renders them a competitive tool in the search for new physics (NP), complementary to both, direct searches at the LHC and Tevatron as well as indirect ones in flavour-changing processes.

As interface between a given theory and experiment typically an effective Hamiltonian is used. The relevant operators are universal and expressed in terms of the light fermion fields and gluons, while their coefficients depend on the details of the theory in question. A model-independent analysis is complicated by the relatively large number of contributing operators and by the fact that the dominant contributions vary for different models. Furthermore, within a given model, in many cases different operators dominate in different regions of the parameter space. Heavy paramagnetic systems are an exception in this respect: their EDMs are dominated by just two terms which are enhanced approximately as $Z^3$; one term is directly proportional to the electron EDM $d_e$, the other stems from electron-nucleon interactions, parametrized by a dimensionless parameter $\tilde C_S$. 

In deriving limits for the electron EDM from the corresponding measurements, commonly
firstly the uncertainties of the numerical proportionality factor are ignored and secondly the other relevant term is set to zero, i.e. it is assumed that no cancellations occur. When performing a quantitative analysis, these issues are obviously important, especially when keeping in mind that theoretical limitations can change the obtained limits by orders of magnitude, as observed for the hadronic limits from the Mercury (Hg) system, see e.g. \cite{Jung:2013xx}. Finally, the obtained limits are usually displayed as ``allowed'' and ``forbidden'' areas in parameter space, making conservative estimates obligatory. 

We address both issues in this paper: the first point is resolved by more careful estimates for the relevant coefficients below. The second issue can be addressed as well, given that at the moment two measurements with similar sensitivities are available, from the Thallium (Tl) and Ytterbiumfluorid (YbF) systems \cite{Regan:2002ta,Hudson:2011zz}. However, the two systems depend on a similar combination of the two terms. Therefore, for the time being, we use in addition a limit from a diamagnetic system, namely Mercury \cite{Griffith:2009zz}. While many terms contribute to that EDM, the one appearing in paramagnetic systems as well is expected to be clearly subdominant; assuming this term to saturate the experimental limit therefore constitutes a conservative estimate. Together, these three systems allow to obtain robust limits for the electron EDM and the coefficient of the electron-nucleon interaction, without the assumption of vanishing cancellations.

The outline for this letter is as follows: the second section is devoted to atom EDMs, with a focus on estimates of the theoretical uncertainties in their relations to $d_e$ and $\tilde C_S$. In Sec.~III, an analogous procedure is carried out for molecules, focusing on the EDM of YbF. The experimental situation is reviewed in Sec.~IV, followed by the phenomenological analysis with present data in Sec.~V, where the new limits on $d_e$ and $\tilde C_S$ are obtained. The results from this analysis allow us to place upper limits on the EDMs of other paramagnetic systems, which we do in Sec.~VI, together with an analysis of the future prospects, before concluding in Sec.~VII.

\section{EDMs of atoms}
For atoms, Schiff's theorem \cite{Schiff:1963zz} implies a vanishing EDM in the non-relativistic limit for systems of particles whose charge distribution is identical to their EDM distribution. The limits from the non-observation of these EDMs are then related to violations of the conditions for this theorem and separated into two classes, depending on which of the approximations is more strongly violated. For reviews, see e.g. Refs.~\cite{Khriplovich:1997ga,Ginges:2003qt}. 

In paramagnetic atoms, which are our main concern, relativistic effects are more important. They are largely enhanced for atoms with a large proton number, scaling at least like $d\sim Z^3$. As mentioned above, this implies a sensitivity mainly to the electron EDM, but also a subset of electron-nucleon interactions is enhanced. The dominant component of the latter is described by
\begin{equation}\label{eq::HeN}
\mathcal{H}_{eN}^S = \frac{G_F}{\sqrt{2}}\sum_{N=n,p}\!\!\tilde{C}_S^N (\bar{N}N)(\bar{e}i\gamma_5e)\,,
\end{equation}
where we neglected operators with other Dirac structures which are negligible here, but have a largely enhanced relative influence in diamagnetic systems. 

In diamagnetic atoms the finite size of the nucleus is the main source for the violation of Schiff's theorem. The dominant contribution to the corresponding EDM stems from its \emph{Schiff moment}, which is in turn related to quark (colour) EDMs and $CP$-violating four-quark interactions. However, the above electron-nucleon interaction is relevant as well. We will use this fact to obtain an upper limit on the coefficient $\tilde C_S$.

\subsection{The EDM of paramagnetic atoms}
For paramagnetic atoms which have one unpaired electron, mainly this electron determines the EDM of the atom, as the effects of the ones in closed shells cancel. Relativistic effects for atoms with large proton number lead to enhancement factors for the electron EDM of $\mathcal{O}(100)$ in these systems. In addition, the coefficient $\tilde{C}_S$ of the electron-nucleon interaction might contribute sizably. Relating the experimentally observable atom EDM to these sources involves complex many-body calculations, for which a number of methods exist (for a review, see e.g. again \cite{Ginges:2003qt}), the results of which sometimes span large ranges. 

The most sensitive system from that class to date is Thallium. Calculations for the enhancement factor yield $d_{\rm Tl}/d_e\sim[-1041,-179]$. One reason for this large range is the presence of strong cancellations between different contributions. 
Recent calculations narrow down this range, however, there is some difference remaining \cite{Dzuba:2009mw,Nataraj:2010vn,Dzuba:2011xx,Nataraj:2011pw,Porsev:2012zx,2012arXiv1202.5402N}. In order to assess its influence, we define two input sets: input set~I is using the result $d_{\rm Tl}\supset -573(20)d_e$ from \cite{Porsev:2012zx}, which includes the value from \cite{Dzuba:2009mw}. Set~II is using the value recommended in \cite{Nataraj:2010vn}, $d_{\rm Tl}\supset -466(10)d_e$, which is about $20\%$ lower. As both calculations claim a precision clearly excluding the other, this situation indicates large systematic effects in one or several of the calculations.

For paramagnetic atoms, the parametrization in Eq.~(\ref{eq::HeN})  
leads in the limit of infinite nucleon mass to 
\begin{equation}
\mathcal{H}_{eN}^S=i G_F/\sqrt{2}\sum_{N=n,p}\tilde{C}_S^NZ_N\gamma_0\gamma_5\rho_N(r)\,,
\end{equation}
with the nuclear densities $\rho_N(r)$ normalized to unity and $Z_N$ denoting the number of the corresponding nucleon in the nucleus. Furthermore assuming $\rho_N(r)\equiv \rho(r)$ and abbreviating\footnote{Note that this definition in principle implies a dependence of $\tilde C_S$ on the system considered. However, because of $(Z_n+Z_p)/A=1$ and $\tilde C_S^n\approx \tilde C_S^p$, this is usually neglected. In addition, the ratios $Z_{p,n}/A$ are approximately universal for all atoms considered here anyway.} $\tilde{C}_S=\sum_N Z_NA^{-1}\tilde{C}_S^N$ leads to 
\begin{equation}\label{Eq::CS}
\mathcal{H}_{eN}^S=i G_F/\sqrt{2}A\,\tilde{C}_S\gamma_0\gamma_5\rho(r)\,,
\end{equation} 
which is the Hamiltonian typically used in the atomic calculations for the corresponding coefficient. It is obtained in the same kinds of calculations like the one for the electron EDM and is plagued by the same cancellations. The most recent results yield $d_{\rm Tl}(\tilde{C}_S)=-7.0(2)\times10^{-18}e\,{\rm cm}\,\tilde{C}_S$ \cite{Dzuba:2009kn} and $d_{\rm Tl}(\tilde{C}_S)=-4.06(2)\times10^{-18}e\,{\rm cm}\,\tilde{C}_S$ \cite{Chaudhuri:2008xx}.
We assign the former to input set~I and the latter to set~II. 
The combinations with the second term read
\begin{eqnarray}\label{eq::TlEDM}
d_{\rm Tl}^I &=& -(573\pm20) d_e-(7.0\pm0.3)\times10^{-18}e\,{\rm cm}\,\tilde C_S,\\
d_{\rm Tl}^{II} &=& -(466\pm16) d_e-(4.1\pm0.1)\times10^{-18}e\,{\rm cm}\,\tilde C_S,
\label{eq::TlEDMII}
\end{eqnarray}
where we increased slightly the uncertainties of those coefficients which have not been confirmed by an independent calculation. 
We note that the values for set~I are consistent with the ratio obtained analytically in \cite{PhysRevA.85.029901}, while the ratio of the values for set~II are significantly higher.

Another interesting system is Cesium (Cs), for which several measurements are prepared at the moment, see Table~\ref{tab::ExpEDM}. For this system, the cancellations commented upon above are absent, leading to a more stable prediction. Recent calculations yield compatible results, $d_{\rm Cs}=(120.5\pm1.3)d_e+(0.801\pm 0.004)\times 10^{-18}\,\tilde{C}_S$ \cite{Chaudhuri:2008xx,PhysRevLett.101.033002} and $d_{\rm Cs}=(124\pm4)d_e+(0.759\pm 0.022)\times 10^{-18}\,\tilde{C}_S\,e\,{\rm cm}$ \cite{Dzuba:2009mw}, motivating
\begin{equation}\label{eq::CsEDM}
d_{\rm Cs} = (123\pm4) d_e+(0.78\pm0.02)\times10^{-18}~e\,{\rm cm}\,\tilde C_S\,, 
\end{equation}
which constitutes in this case an even more conservative estimate. This result is consistent with the ratio obtained in \cite{PhysRevA.85.029901}.

For Rubidium (Rb), the calculations are similarly stable, and a very sensitive measurement is prepared as well, see Table~\ref{tab::ExpEDM}. We obtain \cite{PhysRevLett.101.033002} 
\begin{equation}\label{eq::RbEDM}
d_{\rm Rb} = (25.7\pm0.8) d_e+(0.110\pm0.003)\times10^{-18}~e\,{\rm cm}\,\tilde C_S\,. 
\end{equation}
Note that in this case only one recent calculation exists for the single coefficients. The  uncertainty chosen reflects the difference to the analytic ratio given in \cite{PhysRevA.85.029901} and is of similar size as the largest difference between experimentally and theoretically determined  $CP$-conserving quantities in \cite{PhysRevLett.101.033002}.

Finally, there are also plans to measure the EDM of the heaviest alkali atom, Francium (Fr), see once more Table~\ref{tab::ExpEDM}. For this system, even larger enhancement factors are expected, $d_{\rm Fr}/d_e\sim900$ \cite{PhysRevA.59.3082,doi:10.1021/jp904020s}. The coefficient of the electron-nucleon contribution has not been calculated yet, we use the results of \cite{PhysRevA.85.029901} to estimate its value and add an additional $10\%$ uncertainty for that in light of the level of agreement for the atoms discussed above. Of course a dedicated study of the second coefficient would be welcome to confirm this estimate. The result reads
\begin{equation}\label{eq::FrEDM}
d_{\rm Fr}=(903\pm45)d_e+(10.9\pm1.7)\times10^{-18}e\,{\rm cm}\,\tilde C_S\,,
\end{equation}
where we conservatively assigned the estimated $5\%$ uncertainty in \cite{PhysRevA.59.3082} to the coefficient of $d_e$.

\subsection{The EDM of Mercury}
For diamagnetic atoms, i.e. atoms with vanishing total angular momentum, mainly finite-size effects of the nucleus determine the EDM. More specifically, its dominant source is the $CP$-odd nuclear Schiff moment \cite{Schiff:1963zz}.
However, in the following we will make use of the fact that additional sources from electron-nucleon interactions and the electron EDM are present. Regarding the latter, the value usually used in the literature for Hg reads $d_{\rm Hg}(d_e)=1.16\times10^{-2}d_e$ \cite{1402-4896-36-3-011}. The corresponding calculation, however, shows a high sensitivity to higher order effects; the ``corrections'' to a previous estimate \cite{Flambaum:1985gx} amount to $\sim200\%$ and change the sign. The authors point out the sensitivity to correlation effects (which have been found to be large for Hg for its other coefficients), making a new calculation mandatory. 
In light of this situation we do not see a way to extract a meaningful upper limit on the electron EDM from Hg until the theoretical situation improves. However, even taking the central value quoted above, the bound would be weaker than the one from Tl or YbF.

The electron-nucleon interactions are induced via the operators in $\mathcal{H}_{eN}=\sum_{i=S,P,T}\mathcal{H}_{eN}^i$. 
The coefficients in the expression for $d_{\rm Hg}(\tilde C_{S,P,T})$ are obtained again in atomic calculations; usually only the coefficient of the tensor operator is calculated, defined via $\mathcal{H}_{eN}^T=G_F/\sqrt{2}\sum_N\tilde{C}_T^N(\bar{N}i\gamma_5\sigma^{\mu\nu}N)(\bar{e}\sigma_{\mu\nu}e)$, and analytic relations are used to obtain the others \cite{Flambaum:1985gx,Kozlov:1988qn,Ginges:2003qt,Dzuba:2009kn}.\footnote{Note the different conventions for $d_{\rm atom}^{T,P}$ in different publications, e.g. $\mathbf{d}_{\rm atom}^{T,P}=d_{\rm atom}^{P,T}\langle\boldsymbol{\sigma}_N\rangle$ versus $\mathbf{d}_{\rm atom}^{T,P}=d_{\rm atom}^{T,P}\mathbf{I}/I$.} The one relevant in this context reads
\begin{equation}\label{eq::CSCT}
\tilde C_S\frac{\mathbf{I}}{I}\leftrightarrow 1.9\times10^3\left(1+0.3\,Z^2\alpha^2\right)^{-1}A^{-2/3}\mu^{-1}\times\tilde C_T\langle\boldsymbol{\sigma}_N\rangle\,,
\end{equation}
where $\tilde C_T\langle\boldsymbol{\sigma}_N\rangle=\left(\tilde C_T^p\langle\boldsymbol{\sigma}_p\rangle+\tilde C_T^n\langle\boldsymbol{\sigma}_n\rangle\right)$, $\langle\boldsymbol{\sigma}_{p,n}\rangle$ implies the average over the protons/neutrons in the nuclear state and $\mu$ denotes the magnetic moment of the nucleus in terms of the nuclear magneton $\mu_N$. 
We expect the uncertainty for these relations to be small, $\mathcal{O}(\%)$, and therefore negligible in this context, as also indicated by a recent explicit calculation for a variety of atoms \cite{Dzuba:2009kn}.
For the tensor coefficient, defined by
\begin{equation}
\mathbf{d}_{\rm Hg}(\tilde{C}_T) = C_{C_T}^{\rm Hg}\times10^{-20}\tilde{C}_T\langle\boldsymbol{\sigma}_N\rangle e\,{\rm cm}\,,
\end{equation}
recent results read $C_{C_T}^{\rm Hg}=-5.1$ \cite{Dzuba:2009kn} and $C_{C_T}^{\rm Hg}=-4.3$ \cite{Latha:2009nq}.
Thus we obtain, using Eq.~(\ref{eq::CSCT}),
\begin{equation}\label{eq::HgEDM}
d_{\rm Hg}(\tilde C_S) = 
-(0.00047\pm0.00005) \tilde C_S\times10^{-18}e\,{\rm cm}\,,
\end{equation}
where we inserted $\mu_{\rm Hg}=0.506\,\mu_N$\footnote{Source: WebElements (\texttt http://www.webelements.com/)} and $\langle\boldsymbol{\sigma}_N\rangle=-1/3\,\mathbf{I}/I$, the estimate from a simple shell model for the nucleus, and used the common convention $\mathbf{d}=d\mathbf{I}/I$.
For a more detailed analysis of this system, the reader is referred to \cite{Jung:2013xx}.

\section{The EDM of paramagnetic molecules}
Polar molecules exhibit very large internal fields, which average out to zero in absence of an external field due to molecular rotation. The application of an external field mixes rotational levels of opposite parity and induces two effects: one energy split which is sometimes called somewhat sloppily an EDM, because it scales as $|\mathbf{E}_{\rm ext}|$ for sizable fields, but is T-even, and a much smaller one, which is actually T-odd, in which we are interested and which is described below. 
The main difference to atoms is that the external field is only used to prohibit the cancellation of the effect of the internal field, which is the one acting on the electrons. This is why polar molecules can exhibit huge enhancement factors, increasing the sensitivity to fundamental parameters like $d_e$ \cite{Sandars:1967zz}. Analogously to atoms, the molecules are categorized according to the total angular momentum of their electrons. We discuss in the following the paramagnetic case.

The sensitivity of paramagnetic molecules therefore stems in principle from the same mechanism as in paramagnetic atoms, but is even higher. As in the case of atoms, the two main sources are the electron EDM and electron-nucleon interactions. Different molecules like YbF or PbO are used, which provide a naturally high polarizability.  
They exhibit effective amplification factors of internal versus external fields of $\mathcal{O}(10^6)$, resulting in principle in a sensitivity to the electron EDM of $\mathcal{O}(100)$ times that for atoms. 

From the theory point of view, the difficulty lies in calculating the relevant internal field, $\mathbf{E}_{\rm int}$, which cannot be measured. 
For this, again multi-body calculations are employed, which are 
complicated by the presence of the second core and, as before, the large number of electrons. 
The corresponding interaction energy can be written as
\begin{equation}\label{Eq::EDMYbF}
\Delta E=-\langle\mathbf{d}_{\rm YbF}\cdot \mathbf{E}_{\rm ext}\rangle=
\frac{1}{2}\left(W_d\,d_e+W_c\,\tilde{C}_{S}\right)\langle\hat{\mathbf{n}}\cdot\hat{\mathbf{z}}\rangle(E_{\rm ext})\,,
\end{equation}
with an external electric field $\mathbf{E}_{\rm ext}=E_{\rm ext}\hat{\mathbf{z}}$, $\hat{\mathbf{n}}$ denoting the direction of the molecular axis and their alignment depending on the external field. The factor $1/2$ is due to the spin of the electron\footnote{Note again the presence of different conventions in the literature: $W_d$ is sometimes defined without this factor.} and the constant $\tilde{C}_S$ has been introduced in Eq.~\eqref{Eq::CS}. In \cite{Hudson:2011zz}, $\langle\hat{\mathbf{n}}\cdot\hat{\mathbf{z}}\rangle(E_{\rm ext})=0.558$ 
holds \cite{Hudson:2002az}. 
The constant $W_d/2$ reflects the maximal effective electric field acting on the valence electron.
As noted above, in contrast to the atomic case, the effective electric field is now given in terms of the internal field, the effect of which stops canceling out once the external field is applied, due to the closeness of the corresponding rotational levels.

Again in parallel to the experimental efforts there has been recent theory activity. The relevant results for YbF are shown in Table~\ref{tab::YbF}. As pointed out in \cite{PhysRevA.85.029901}, the ratios of these matrix elements can be estimated analytically. 
Their value for YbF, $W_d/W_c=114\times 10^{18}/e\,{\rm cm}$, is in agreement with the latest calculations \cite{PhysRevA.78.012506,1742-6596-80-1-012051,Chaudhuri:2009} within $\sim 10\%$\footnote{Note that Ref.~\cite{PhysRevA.78.010502} is aiming at an analytical estimate rather than high precision, they estimate the accuracy to $\sim25\%$, making their result compatible with the following estimate.}, reflecting the spread in the values for $W_d$. We conservatively allow for these $10\%$ variation in both directions as an error estimate. In absence of a second recent determination of $W_c$, we assign it as well as an error estimate there, which yields finally
\begin{equation}\label{Eq::Wdc}
W_d=-(1.1\pm 0.1)\times 10^{25}\,{\rm Hz}/e\,{\rm cm}\,,\quad W_c=-(92\pm 9)\,{\rm kHz}\,.
\end{equation}
%
%
\begin{table}
\begin{center}
\begin{tabular}{cccc}\hline\hline
$W_d(10^{25}{\rm Hz}/e\,{\rm cm})$ & $W_c({\rm kHz})$ & Ref. &  Year\\\hline
-0.91 & -\phantom{1}82 & \cite{PhysRevLett.77.5346} & 1996\\
-1.26 & -120 & \cite{0953-4075-30-18-003,PhysRevA.49.4502} & 1994/97\\
-1.20 & -104 & \cite{0953-4075-31-3-003} &  1998\\
-1.20 & -108 & \cite{0953-4075-31-7-008} &  1998\\
-1.21 & -- & \cite{mosyagin-1998-31} &  1998\\
-1.50 & -- & \cite{PhysRevA.78.010502} &  2008\\
-1.04 & -92 & \cite{PhysRevA.78.012506,1742-6596-80-1-012051} &  2007/08\\
-1.16 &-- & \cite{Chaudhuri:2009} &  2009\\\hline\hline
\end{tabular}
\end{center}
\caption{\label{tab::YbF}Calculations for the coefficients in the expression for the dipole moment of YbF.} 
\end{table}%
We note that a calculation of $W_c$ by a second group with the presently available methods would be welcome. From these considerations we finally obtain
\begin{equation}\label{eq::YbFEDM}
d_{\rm YbF} = -(1.3\pm0.1)\times10^6 d_e-(9740\pm960)\times10^{-18}~e\,{\rm cm}\,\tilde C_S\,,
\end{equation}
to be compared with Eqs.~(\ref{eq::TlEDM})-(\ref{eq::FrEDM}).

\section{Experimental status}
At present, the most stringent limits relevant to the extraction of $d_e$ and $\tilde C_S$ stem from searches for EDMs of Tl \cite{Regan:2002ta}, YbF \cite{Hudson:2011zz} and Hg \cite{Griffith:2009zz}, see Table~\ref{tab::ExpEDM}. Although these limits have different orders of magnitude, their different dependence on the fundamental parameters of the theory actually leads to similar sensitivities. Especially, despite the very different factors in Eqs.~\eqref{eq::TlEDM}/\eqref{eq::TlEDMII} and (\ref{eq::YbFEDM}), the resulting limits for the electron EDM are similar so far.
Note, however, that the result for YbF is still statistically limited.

\begin{table}
\begin{center}
\begin{tabular}{lcc}\hline\hline
System & Present limit $(e~{\rm cm})$ & Expected limit $(e~{\rm cm})$  \\\hline
$\phantom{}^{199}$Hg  &
$(0.49\pm1.50)\times 10^{-29}$ \cite{Griffith:2009zz}    & ---\\
$\phantom{}^{205}$Tl  & $-(4.0\pm 4.3)\times 10^{-25}$ \cite{Regan:2002ta}       & ---\\
${\rm \phantom{}^{133}Cs}^{*}$  & $1.4\times 10^{-23}$ \cite{PhysRevLett.63.965} & $\mathcal{O}(10^{-26}/10^{-27})$ \cite{Amini:2007ku,2004APS..DMP.P1056K,Weissetal}\\
$\phantom{}^{\phantom{1}85}$Rb   & $1\times10^{-18}$ \cite{PhysRev.153.36}    & $\mathcal{O}(10^{-27}/10^{-28})$ \cite{Weissetal} \\
& $(1.2\times10^{-23})^\dagger$ \cite{RbEDMthesis} &\\
$\phantom{}^{210}$Fr  & ---                                                 & $\mathcal{O}(10^{-26}/10^{-29})$ \cite{Sakemi:2011zz,PhysRevX.2.041009}\\
$\phantom{}^{\phantom{210}}$YbF                 & $(3.5\pm8.7)\times 10^{-22}$ \cite{Hudson:2011zz,Kara:2012ay} & $\mathcal{O}(10^{-22}/10^{-23-24})$ \cite{Kara:2012ay} \\
\hline\hline
\end{tabular}
\caption{\label{tab::ExpEDM}Present limits on absolute values of EDMs at $95\%$~CL for the most sensitive atoms/molecules, together with short-term/mid-term expected sensitivities.
${}^{*}$: Given in the paper as $(-1.8\pm 6.7\pm1.8)\times10^{-24}e\,{\rm cm}$. ${}^\dagger$: unpublished}
\end{center}
\end{table}

Recently there have been several developments which allow to expect significantly improved sensitivities in the near future, see also e.g. \cite{Ginges:2003qt,Pospelov:2005pr,Raidal:2008jk,Fukuyama:2012np}: the first option is to improve the methods described above.
With the experiments for Tl completely dominated by their systematic errors, significant advancement seems difficult within this system. An improvement, up to two orders of magnitude, might come instead from the Cs, Rb and Fr systems \cite{Amini:2007ku,2004APS..DMP.P1056K,Weissetal,Sakemi:2011zz,PhysRevX.2.041009}. The expected limits correspond to probing the electron EDM to  $\lesssim 10^{-29}e~{\rm cm}$ in the short-term future (1-3 years), and even sensitivities down to $10^{-31}e\,{\rm cm}$ seem achievable.

Further measurements with paramagnetic molecules are expected to strengthen the present limit by another order of magnitude or more for YbF, and many more systems are explored as well, see e.g. \cite{Fukuyama:2012np} for a recent list, making for an expected improvement of at least the one from atoms. 

In the future, trapped molecular ions might also be used as sensitive probes for EDMs, however, at the moment there are still severe experimental and theoretical challenges to overcome. Furthermore, also solid state systems are being explored as sensitive probes for the electron EDM \cite{Lamoreaux:2001hb,0038-5670-11-3-A11}. While again some experimental as well as theoretical progress is necessary before competitive results can be achieved, recent results show the progress in this field \cite{Eckel:2012aw}.
Finally, new techniques are being explored for measuring the EDMs of charged particles directly by using a storage ring \cite{Garwin:1959xx,Semertzidis:1998sp,Khriplovich:1998zq,Farley:2003wt}. While the main focus here is on other systems, 
there are also proposals to use the technique for molecular ions, see e.g. \cite{Kawall:2011zz}.

The plethora of ongoing and planned experiments, all aiming at the strengthening of present limits by several orders of magnitude, will take this field to a new level. Especially if one or several of these experiments should result in a significant non-zero signal, the question of a more refined analysis of the various uncertainties will be posed, making a global analysis obligatory. We will explore steps in this direction below.

\section{A robust limit on the electron EDM\label{sec::eEDM}}
With the results of the last sections at hand, we proceed to derive limits on the electron EDM and the electron-nucleon coupling. We do this in two steps: first, we derive the limit just from the measurements with Tl and YbF, to avoid even input from the conservative bound on $\tilde C_S$ from Hg. Then we add this as a third constraint, obtaining a much stronger limit on both, $d_e$ and $\tilde C_S$. We use the data as given in Table~\ref{tab::ExpEDM}, i.e. not (yet) transforming the given values into symmetric bounds. 

\begin{figure}[tb]
\begin{center}
\includegraphics[width=6.6cm]{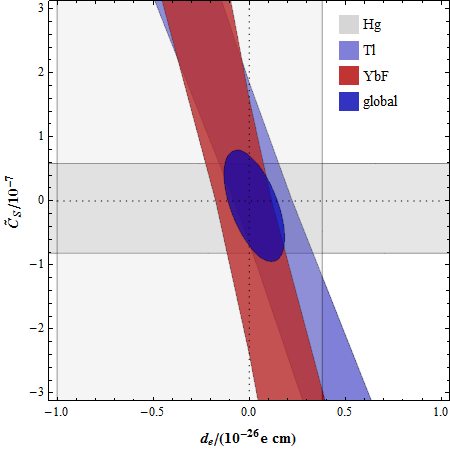}
\end{center}
\caption{\label{fig::deCStilde} Bounds from Hg, Tl~and~YbF in the $d_e$--$\tilde C_S$-plane, using input set~I. The very light grey vertical bound indicates the 1D-limit on $d_e$ when using only the Tl and YbF constraints without the aid of Hg.} 
\end{figure}

\begin{figure}[tb]
\begin{center}
\includegraphics[width=6.6cm]{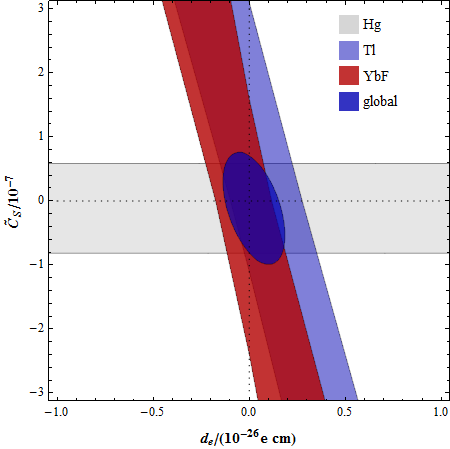}
\end{center}
\caption{\label{fig::deCStildeII} Bounds from Hg, Tl~and~YbF in the $d_e$--$\tilde C_S$-plane, using input set~II.} 
\end{figure}

The results are shown in Figs.~\ref{fig::deCStilde} and~\ref{fig::deCStildeII} for the two input sets, where the constraint from each system is shown in the $d_e-\tilde C_S$--plane. We illustrate by the light grey area in Fig.~\ref{fig::deCStilde} the bound on the electron EDM obtained by the combination of the Tl and YbF constraints only (compare also to \cite{PhysRevA.85.029901}). For input set~II, these two constraints are basically parallel; therefore no bound can be obtained in this case without further assumptions. 
The dark area in the middle is the global fit to all three constraints. The projections on the parameters of interest read
\begin{eqnarray}\label{eq::deresult}
d_e^I&=&(0.026\pm0.065)\times10^{-26}e\,\rm{cm}\quad\mbox{and}\\
\tilde C_S^I&=&(-0.08\pm0.36)\times10^{-7}\label{eq::CSresult}
\end{eqnarray}
for input set~I and
\begin{eqnarray}\label{eq::deresultII}
d_e^{II}&=&(0.024\pm0.066)\times10^{-26}e\,\rm{cm}\quad\mbox{and}\\
\tilde C_S^{II}&=&(-0.12\pm0.36)\times10^{-7}\label{eq::CSresultII}
\end{eqnarray}
for input set~II. These values are
to be compared with $d_e=(-0.31\pm0.35)\times10^{-26}e\,{\rm cm}$ and $\tilde C_S=(3.2\pm3.3)\times10^{-7}$, obtained using only the two constraints from Tl (input set~I) and YbF. The  corresponding upper limits at $95\%$~C.L. are 
\begin{eqnarray}
|d_e^{I,II}|&\leq& 0.14\times 10^{-26}e\,\rm{cm}\quad\mbox{and}\\
|\tilde C_S^{I,II}|&\leq& 0.72(0.74)\times10^{-7}
\end{eqnarray}
for the global fit, whereas $|d_e|\leq 0.89\times 10^{-26}e\,\rm{cm}$ and $|C_S|\leq 8.6\times10^{-7}$ when excluding the input from Hg. Independent of the input set, the global fit therefore results in a limit on the electron EDM very similar to the one obtained naively from YbF alone, but is obtained in a more conservative manner. Using only paramagnetic systems at the moment worsens this limit approximately by a factor of six for input set~I and is not possible for set~II.

In the next section we will show, however, that even the conservative assumption entering here via the input from the Hg system can be avoided with future data.

\section{Upper limits for other systems and future prospects}
The already available limits for the EDMs of Cs and Rb given in Table~\ref{tab::ExpEDM} do not strengthen the limits on $d_e$ and $\tilde C_S$ derived in the last section. This in turn implies that we can place non-trivial bounds on these EDMs from our results in Eqs.(\ref{eq::deresult}),(\ref{eq::CSresult}). To do so, we map the $95\%$~C.L. area from the global fit onto the corresponding interval of the atom EDMs, taking additionally the theoretical uncertainties there into account. As the obtained intervals for the two input sets are very similar, we do not differentiate in the following between the two.

Starting with Cs, we obtain with aid of \refEqws{eq::CsEDM} the $95\%$~C.L. interval
\begin{equation}
d_{\rm Cs}\in[-1.6,2.0]\times10^{-25}e\,{\rm cm}\,.
\end{equation}
Therefore the dedicated experiments are expected to improve the present limit by approximately two orders of magnitude before becoming sensitive to possible non-vanishing contributions. The same is true for the Rb experiments, where the interval reads
\begin{equation}
d_{\rm Rb}\in[-3.4,4.1]\times10^{-26}e\,{\rm cm}\,.
\end{equation}
Also for Fr we obtain a rather strong limit already,
\begin{equation}
d_{\rm Fr}\in[-1.3,1.5]\times10^{-24}e\,{\rm cm}\,.
\end{equation}
However, with the expected final sensitivities, see Table~\ref{tab::ExpEDM}, the planned experiments will be able to improve greatly the present bounds or to finally obtain a non-zero result. A contradicting measurement in one of these systems would indicate a severe issue in one of the involved experiments or the theoretical description.

While the use of the constraint from the Hg system proves very advantageous at the moment, in principle it would be preferable to perform a similar procedure without this input. We therefore investigate to what extend this is possible with coming measurements, see once more Table~\ref{tab::ExpEDM}.
\begin{figure}[tb]
\begin{center}
\includegraphics[width=6.6cm]{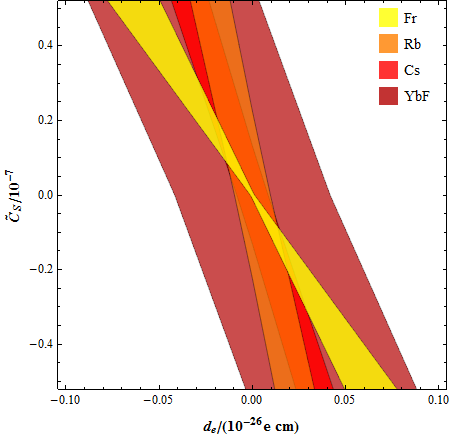}
\end{center}
\caption{\label{fig::deCStildefut1} Bounds from various paramagnetic systems as expected in the short-term future (1-3 years) in the $d_e$--$\tilde C_S$-plane, see Table~\ref{tab::ExpEDM}. Note the different scales compared to Figs.~\ref{fig::deCStilde} and~\ref{fig::deCStildeII}. 
} 
\end{figure}
To that aim we plot in Figs.~\ref{fig::deCStildefut1} and~\ref{fig::deCStildefut2} the constraints expected from the future measurements within 1-3 years and in the longer run, respectively. Note that the plotted areas correspond to $1/60$ ($1/2000$) that of Fig.~\ref{fig::deCStilde}. 
\begin{figure}[tb]
\begin{center}
\includegraphics[width=6.6cm]{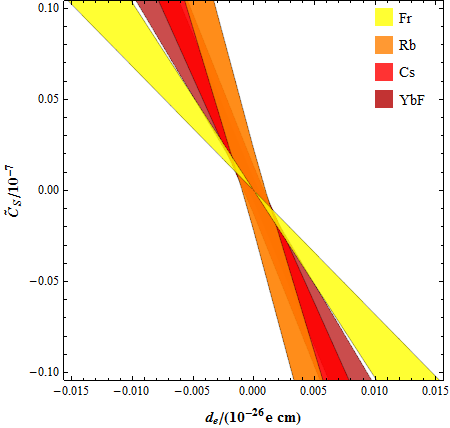}
\end{center}
\caption{\label{fig::deCStildefut2} Bounds from various paramagnetic systems as expected in the mid-term future (more than 3 years) in the $d_e$--$\tilde C_S$-plane, see again Table~\ref{tab::ExpEDM}. Note the different scales compared to Figs.~\ref{fig::deCStilde}-\ref{fig::deCStildefut1}.} 
\end{figure}
The constraints are chosen such that their central value is zero, thereby reflecting the resulting limits in the absence of a non-zero result; for significant measurements, of course all constraints should still overlap. Already in the scenario for the first plot the additional constraint from Hg is rendered unnecessary and the results for $d_e$ and $\tilde C_S$ are significantly improved. 
These plots illustrate clearly the importance of various experiments with different atoms and/or molecules. First of all, at least two competitive measurements are necessary to yield a model-independent constraint on $d_e$. Ideally they should constrain very different combinations of $d_e$ and $\tilde C_S$, as for example Rb and Fr. 
Secondly it is important to have more than two constraints in order to confirm the theoretical description and safeguard against possible systematic issues. Finally, the combination of more constraints yields additional precision, which can indicate non-vanishing values for $d_e$ and $\tilde C_S$ earlier. The list in Table~\ref{tab::ExpEDM} indicates that this challenge is met.

\section{Conclusions}
Measurements of EDMs are extremely sensitive probes of $CP$-violating phases beyond the SM. They therefore have the potential to reveal NP and will continue to strongly constrain possible NP scenarios. The experiments presently planned and constructed will take this field to a new level of precision, challenging many models. To meet that precision, bounds from these measurements should be derived carefully. We have shown in this letter that it is possible to go beyond the common assumption of vanishing cancellations, already with present data. Doing so, we provided expressions for various systems of experimental interest, where we focused on a careful estimate of theory uncertainties. This allowed us to obtain more robust limits on the electron EDM and the electron-nucleon interaction. Despite the more conservative extraction, these limits match the more naively extracted ones in precision, due to the combination of various measurements. At the moment the additional input from the Hg system is necessary, which is possible with conservative assumptions. In the future, even these assumptions can be avoided, once strong limits and/or determinations of the EDMs from more paramagnetic systems are available. 

\section*{Acknowledgements}
The author would like to thank Toni Pich for collaboration on related topics and Bhanu Das, Vladimir Dzuba, Dilip Kumar Singh and Celal Harabati for helpful discussions. 
This work is funded by the German Federal Ministry of Education and Research (BMBF).

\bibliography{EDM}
\end{document}